\documentclass[aps,prd,twocolumn,showpacs,floatfix,preprintnumbers,amsmath,amssymb,nofootinbib,groupedaddress]{revtex4-1}

\input epsf
\usepackage{graphicx}
\usepackage{color}
\usepackage{mathtools}

\newcommand{\beq}{\begin{equation}}
\newcommand{\eeq}{\end{equation}}
\newcommand{\barr}{\begin{eqnarray}}
\newcommand{\earr}{\end{eqnarray}}

\newcommand{\rme}{\textrm{e}}

\newcommand{\bs}{\boldsymbol}
\newcommand{\bk}{\boldsymbol{k}}
\newcommand{\bq}{\boldsymbol{q}}
\newcommand{\bx}{\boldsymbol{x}}

\newcommand{\aap}{Astron. Astrophys.}

\newcommand{\mnras}{Mon. Not. R. Astron. Soc.}

\newcommand{\physrep}{Phys. Rep.}

\newcommand{\jcap}{J. Cosm. Astropart. Phys.}

\newcommand{\lsim}{\mathrel{\hbox{\rlap{\lower.55ex\hbox{$\sim$}} \kern-.3em \raise.4ex \hbox{$<$}}}}
\newcommand{\gsim}{\mathrel{\hbox{\rlap{\lower.55ex\hbox{$\sim$}} \kern-.3em \raise.4ex \hbox{$>$}}}}

\newcommand{\revision}[1]{#1}

\begin{document}
\title{Perturbative interaction approach to cosmological structure formation}
\author{Yacine Ali-Ha\"imoud}
\affiliation{Department of Physics and
  Astronomy, Johns Hopkins University, Baltimore, Maryland 21218, USA}
\date{\today}

\begin{abstract}
A new approach to cosmological perturbation theory has been recently introduced by Bartelmann \emph{et al.}, relying on nonequilibrium statistical theory of classical particles, and treating the gravitational interaction perturbatively. They compute analytic expressions for the nonlinear matter power spectrum, to first order in the interaction, and at one-loop order in the linear power spectrum. The resulting power spectrum is well behaved even at large wavenumbers and seems in good agreement with results from numerical simulations. In this paper, we rederive their results concisely with a different approach, starting from the implicit integral solution to particle trajectories. We derive the matter power spectrum to first order in the interaction, but to arbitrary order in the linear power spectrum, from which the one-loop result follows. We also show that standard linear perturbation theory can only be recovered at infinite order in the gravitational interaction. At finite order in the interaction, we find that the linear power spectrum is systematically and significantly underestimated. A comprehensive study of the convergence of the theory with the order of the interaction for nonlinear scales will be the subject of future work.

\end{abstract}

\maketitle

\section{Introduction}

The statistical properties of cold dark matter (CDM) in the nonlinear regime make for a technically challenging problem, and their study has been the bread and butter of several generations of cosmologists. Various approaches have been pursued to obtain analytic predictions for the matter power spectrum and higher-order statistics beyond the well-established linear perturbation theory results \cite{Ma_1995}. The reader may refer for example to Ref.~\citep{Bernardeau_2002} for a detailed review and to Ref.~\cite{Alvarez_2014} for a list or more recent works.

The most standard analytic approaches to large-scale structure rely on an expansion in variables measuring departures from perfect homogeneity. The perturbation variables are the density and velocity field in the case of Eulerian perturbation theory (see e.g.~Ref.~\cite{Jain_1994}), or the displacement field in the case of Lagrangian perturbation theory (LPT, see e.g.~Ref.~\cite{Bouchet_1995}). In a given gauge, all perturbation schemes agree on a unique linear density field, which matches observations on large scales, on which the departures from homogeneity are indeed small. On quasilinear scales, perturbative approaches improve upon linear theory, but all methods eventually fail dramatically at small enough scales, where inhomogeneities cannot be treated perturbatively \cite{Carlson_2009}.

In a series of recent papers \cite{Bart2, Bart3, Bart4}, Bartelmann \emph{et al.}~(hereafter BFB) have proposed a
new approach, based on non-equilibrium statistical  theory of classical particles and using tools similar to the path-integral approach of quantum field theory. Their approach introduces a new expansion parameter, in addition to (and independently of) the usual perturbation variables\footnote{In BFB the expansion with respect to the usual perturbation variables is carried out when evaluating the initial phase-space probability distribution. The latter is computed perturbatively in the small initial correlations between particles' positions and momenta, which are proportional to the initial matter power spectrum. See Appendix A of Ref.~\citep{Bart2}.}: the order in the gravitational interaction. To set our lexicon right away, throughout this paper we will use ``linear", ``quadratic", ``one-loop" or ``non-linear" to refer to the dependence on standard perturbation variables. When discussing the interaction expansion, we shall say explicitly ``to $n$-th order in the interaction". BFB derive analytic formulae for the nonlinear power spectrum to first order in the interaction, which are well behaved at large wavenumbers and seem to match results from numerical simulations deep into the non-linear regime, way beyond the reach of standard perturbation schemes. 

In this paper we rederive all of BFB's results with a completely different approach, starting from the implicit integral solution for particle trajectories. We obtain explicit solutions for the trajectories or equivalently for the CDM phase-space density, perturbatively in the gravitational interaction. The density field and its power spectrum are then explicitly computed to first order in the gravitational interaction. Our solutions are on the other hand non-perturbative in the standard perturbation variables (or in the initial departure from inhomogeneity), therefore going beyond BFB's results, which are restricted to one-loop order in the linear power spectrum. \revision{From these nonperturbative solutions, we show how BFB obtain their approximate solutions. We argue that they rely on an improper treatment of the ``damping factor", which leads to a spurious enhancement of power at small scales, mimicking the effect of nonlinear growth.}

We also make contact with standard linear perturbation theory. We show that it can only be recovered at infinite order in the gravitational interaction, and with a rather slow rate of convergence as a function of the order in the interaction. This warrants a careful study of the convergence of the theory in the nonlinear regime, where explicit calculations so far (including in the present work) have been limited to first order in the interaction.

This paper is organized as follows. In Section \ref{sec:Zeldovich} we recall standard results about clustering in the Zeldovich approximation, from which the results of Ref.~\cite{Bart3} are immediately derived. In Section \ref{sec:general} we define the new expansion scheme and rederive the main results of Ref.~\cite{Bart4} starting from particle trajectories. Section \ref{sec:shortcomings} makes the connection with standard linear perturbation theory. We conclude in Section \ref{sec:concl}. 

\section{Dark matter statistics in the Zeldovich approximation}\label{sec:Zeldovich}

\subsection{Summary of known results}

In this section we re-derive the results of Ref.~\cite{Bart3}, where the statistics of CDM in the Zeldovich approximation are computed approximately.

The well-known Zeldovich approximation is the first order of LPT \citep{Bernardeau_2002}. The statistics of CDM in this approximation have been extensively studied, see for example Refs.~\citep{Schneider_1995, Tassev_2014}.

In this approximation particle trajectories are given by
\beq
\bs{x}_{\rm Zel}(\bq, \eta) = \bq + \bs{\psi}(\bq, \eta),
\eeq
where $\bx$ is the comoving coordinate, $\eta$ is the conformal time, $\bq$ is the unperturbed Lagrangian coordinate and $\bs{\psi}$ is the displacement field. The latter is a curl-free gaussian random field. It is related to the linear density field $\delta_{\rm L}$ through $\bs \nabla \cdot \bs \psi(\bx, \eta) = - \delta_{\rm L}(\bx, \eta)$, or equivalently, in Fourier\footnote{Our Fourier transform convention is \\$f(\bk) = \int d^3 x~ \rme^{i \bk \cdot \bx} f(\bx), ~~ f(\bx) = \int d^3k/(2 \pi)^3 ~\rme^{-i \bk \cdot \bx} f(\bk)$.} space
\beq
\bs\psi(\bk, \eta) = -i \frac{\bk}{k^2} \delta_{\rm L}(\bk, \eta).\label{eq:psi-delta-fourier}
\eeq
On subhorizon scales, the linear density field grows with a scale-independent growth factor:
\beq
\delta_{\rm L}(\bk, \eta) = D(\eta) \delta_{\rm L}(\bk, \eta_i),
\eeq
where $\eta_i$ is some initial time. This implies that the displacement field remains parallel to itself and that trajectories are straight lines. In particular, the peculiar velocity field is 
\beq
\frac{d \bx_{\rm Zel}}{d \eta} = \frac{\dot{D}}{D} \bs{\psi}(\bq, \eta).
\eeq

The Zeldovich approximation is only accurate to linear order in the density field, but one can still compute the clustering statistics that would result if particles followed these trajectories. The overdensity field $\delta(\bx)$ is then obtained by summing over the streams leading to the position $\bx$:
\beq
1 + \delta_{\rm Zel}(\bx, \eta) = \int d^3 q ~\delta_{\rm D}\left(\bx - \bx_{\rm Zel}(\bq, \eta)\right),
\eeq
where $\delta_{\rm D}$ is the Dirac distribution. Taking the Fourier transform, we get \cite{Crocce_2006}
\beq
\delta_{\rm Zel}(\bk) = \int d^3 q ~ \rme^{i \bk \cdot \bq}\left[\rme^{ i \bk \cdot \bs{\psi}(\bq)}  - 1\right]. \label{eq:delta-Zel}
\eeq
The power spectrum [defined by $\langle \delta(\bk) \delta^*(\bk')\rangle = (2 \pi)^3 P(k) \delta_{\rm D}(\bk)$] follows immediately \cite{Crocce_2006}:
\beq
P_{\rm Zel}(k) = \rme^{- k^2 \sigma_{\psi}^2}\int d^3 q ~\rme^{i \bs k \cdot \bq} \left[\rme^{k^i k^j C_{ij}(\bq)} -1 \right], \label{eq:Pk-Zel} 
\eeq
where $C_{ij}(\bq) \equiv \langle \psi_i(\bs{0}) \psi_j(\bq)\rangle$ is the correlation tensor of the displacement field and is given by
\beq
C_{ij}(\bq)= \int \frac{d^3 k}{(2 \pi)^3} \rme^{-i \bk \cdot \bq} \frac{k^i k^j}{k^4}P_{\rm L}(k). \label{eq:Cij}
\eeq
In equation \eqref{eq:Cij}, $P_{\rm L}$ is the power spectrum of the linear density field, and $\sigma_{\psi}^2$ is the variance of the displacement field (per axis). It is obtained from 
\beq
\sigma_{\psi}^2 = \frac13 \int \frac{d^3 k}{(2 \pi)^3}  \frac{P_{\rm L}(k)}{k^2}.
\eeq
At zero separation, we have $C_{ij}(\bs 0) = \sigma_{\psi}^2 \delta_{ij}$. 

\begin{figure}
\includegraphics[width = 85 mm]{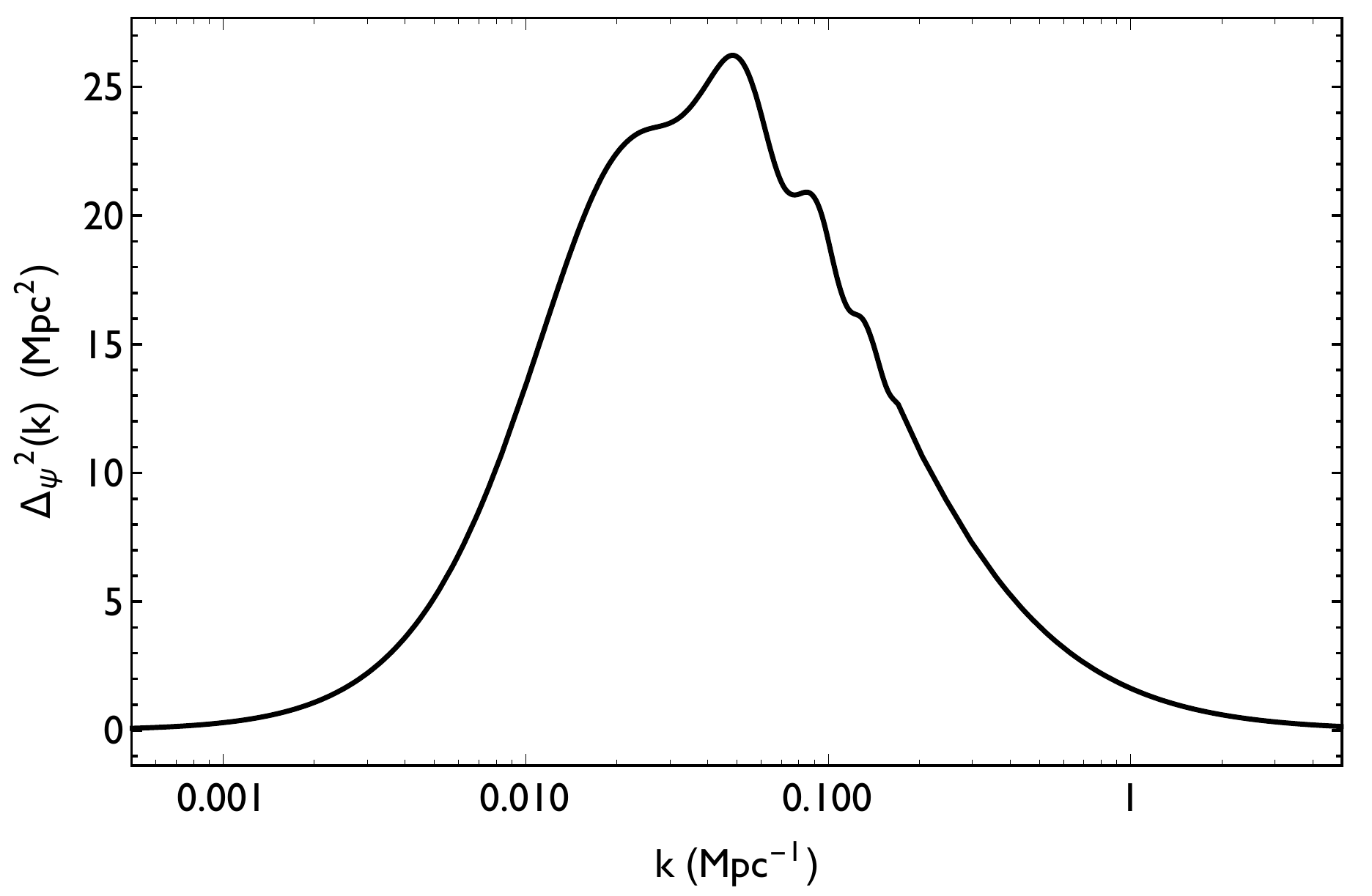}
\caption{Variance of the displacement field per axis, per logarithmic $k$-interval, at redshift zero: $\Delta_{\psi}^2 \equiv \frac13 k P_{\rm L}(k)/(2 \pi^2)$. The linear power spectrum was obtained with \textsc{Camb} \cite{Lewis:1999bs} for cosmological parameters consistent with \emph{Planck} results \cite{Planck_2014}.} \label{fig:psi}
\end{figure}

We show the characteristic variance of the displacement field per axis, as a function of scale, in Fig.~\ref{fig:psi}. We see that the Zeldovich displacement field fluctuates mostly on scales $k \lesssim 1$ Mpc$^{-1}$.

The main advantage of the Zeldovich approximation over linear theory is its better correlation with the fully nonlinear density field \cite{Tassev_2012}. This arises because the Zeldovich approximation accounts for the advection of structure by large-scale flows, which cannot be accounted for by standard linear perturbation theory \cite{Tassev_2012}.

It is however well known that the nonlinear power spectrum in the Zeldovich approximation is actually lower than the linear prediction \cite{Crocce_2006}, \revision{and a fortiori lower than the nonlinear power-spectrum. Indeed, in this approximation, structure is exponentially suppressed on scales smaller than the characteristic displacement, $\sigma_{\psi} \approx 9$ Mpc at redshift 0 \cite{Tassev_2012}, i.e.~for wavenumbers $k \gtrsim \sigma_{\psi}^{-1} \approx 0.1$ Mpc$^{-1}$. This is a consequence of the unphysical assumption that particles move forever on straight lines, even after streams cross. Actual particle trajectories are strongly deflected at stream crossing, and CDM eventually forms bound structures. To account for this requires a framework to treat nonlinear evolution, using for example higher-order analytic methods or numerical simulations.}

We now Taylor-expand equation \eqref{eq:Pk-Zel} to second order in $P_{\rm L}$ (recalling that $\sigma_{\psi}^2$ is of the same order as $P_{\rm L}$), and arrive at \cite{Crocce_2006}
\beq
P_{\rm Zel}(k) \approx P_{\rm L}(k) - k^2 \sigma_{\psi}^2 P_{\rm L}(k) + P_{\textrm{1-loop}}(k), \label{eq:Pzel-taylor}
\eeq
where
\barr
P_{\textrm{1-loop}}(k) &\equiv& \frac{1}{2} \int \frac{d^3
  k'}{(2 \pi)^3} P_L(\bk') P_L(\bs{k} - \bs{k}')  \nonumber\\
 &&~~~ \times \left(\frac{\bs{k}
 \cdot \bs{k}'}{(\bk')^2}\right)^2 \left(\frac{\bs{k}
 \cdot (\bs{k} - \bs{k}')}{(\bs{k} - \bs{k}')^2}\right)^2.~~~~~\label{eq:P-1loop}
\earr
\revision{As expected (and by construction!), the linear term is just the linear power spectrum. The quadratic terms arise from the nonlinear mapping between the density field and the displacement field. The term $- k^2 \sigma_{\psi}^2 P_{\rm L}(k)$ cancels out the contributions of the long-wavelength modes $k' \ll k$ of the loop integral to the power spectrum $P(k)$. Physically, it embodies the fact that as long-wavelength modes of the displacement field cannot have an effect on the small-scale power spectrum \cite{Senatore_2014}. 
}  

\revision{Since the Zeldovich approximation is only the first order of LPT, the resulting power spectrum is only meaningful and accurate to first order in the linear power spectrum, and should only be trusted for $k \lesssim \sigma_{\psi}^{-1} \approx 0.1$ Mpc$^{-1}$. Obtaining the correct term of order $P_{\rm L}^2$ in the CDM power spectrum would require going to third order in perturbation theory.
}

\revision{\subsection{Discussion of BFB's result}}

\revision{In Ref.~\citep{Bart3}, BFB derive the expected clustering of particles treated with the ``Zeldovich propagator", i.e.~enforcing that their trajectories are those given by the Zeldovich approximation. Without additional physical ingredients to deal with the deflection of trajectories at stream crossing, their treatment should (but does not, as we shall see) recover the known Zeldovich power spectrum, Eq.~\eqref{eq:Pk-Zel}.}

\revision{The statistical treatment of BFB requires them to compute the joint probability distribution of the positions and momenta of $N$ particles at some early initial time. They partially expand this probability distribution, to second order in the position-position and position-momentum correlations, but keeping the full exponential dependence on the correlation matrix of momenta (Sections IV E and IV F of Ref.~\citep{Bart2}). As a result, they arrive at a partially expanded power spectrum for the Zeldovich approximation (Eqs.~(44) and (48) of Ref.~\citep{Bart3}). With our notation\footnote{The notation correspondence is $\sigma_1^2/3 \equiv \sigma_{\psi}^2$,  $1+\tau_1 \equiv D(\eta)/D(\eta_i)$. Ref.~\cite{Bart3} finds a quadratic contribution proportional to $\tau_1^4$ instead of $(1 + \tau_1)^4$. This difference arises from the choice of initial conditions: a gaussian density field for BFB, and a gaussian displacement field in our case. It is unimportant at late times.}, it takes the form 
\beq
P_{\rm Zel, BFB}(k) = \rme^{- k^2 \sigma_{\psi}^2} \left[P_{\rm L}(k) +  P_{\textrm{1-loop}}(k)\right],
\eeq
where the 1-loop term is given by our equation \eqref{eq:P-1loop}. At this point, BFB correctly remark (Section V B of Ref.~\cite{Bart3}) that since the term in brackets only accounts for correlations of the initial phase-space distribution to second order, it is necessary to also approximate the exponential damping prefactor at the same order. However, instead of writing a consistent Taylor expansion, which would lead to our Eq.~\eqref{eq:Pzel-taylor}, they approximate the damping term $\rme^{- k^2 \sigma_{\psi}^2}$ by $(1 + k^2 \sigma_{\psi}^2)^{-1}$, and apply it \emph{only} to the one-loop term, so their final result is 
\beq
P_{\rm Zel, BFB, final}(k) = P_{\rm
    L}(k) + \frac{P_{\textrm{1-loop}}(k)}{ 1+ k^2 \sigma_{\psi}^2}. \label{eq:PZel-Bart}
\eeq 
Their justification of this choice is that the exponential suppression in the Zeldovich approximation is unphysical, and should therefore not be applied to the linear term. The choice of replacing the negative exponential by a fraction is not discussed.  
}

For $k \gtrsim 1$ Mpc$^{-1}$, the loop integral is converged when including only $k' \ll k$ and $|\bk' - \bk| \ll k$ contributions (see Fig.~\ref{fig:psi}), and is approximately 
\beq
P_{\textrm{1-loop}}(k \gtrsim 1 \textrm{ Mpc}^{-1}) \approx k^2 \sigma_{\psi}^2 P_{\rm L}(k).
\eeq 
\revision{This implies that equation Eq.~\eqref{eq:PZel-Bart} has the asymptotic value 
\beq
P_{\rm Zel, BFB, final}(k \gtrsim 1 \textrm{ Mpc}^{-1}) \approx 2 \times P_{\rm L}(k),
\eeq
which is precisely the behavior seen in Fig.~1 of Ref.~\citep{Bart3}.
}

\revision{Whereas the resulting power spectrum \emph{appears to be} closer to a nonlinear power spectrum than what one would obtain from the Zeldovich approximation (it is enhanced instead of exponentially damped at small scales), arriving at Equation (\ref{eq:PZel-Bart}) results from a series of choices that are neither mathematically consistent nor physically justified.}

\subsection{Bispectrum}

We now compute the bispectrum $B(\bk_1, \bk_2, \bk_3)$ [defined as usual by $\langle \delta(\bk_1) \delta(\bk_2) \delta(\bk_3)\rangle = (2 \pi)^3 \delta_{\rm D}(\bk_1 + \bk_2 + \bk_3) B(\bk_1, \bk_2, \bk_3)$] in the Zeldovich approximation. We recall that the Zeldovich approximation is only the first order of LPT and therefore \emph{is not of high enough order in standard perturbation variables to compute the bispectrum}. One should indeed go to second order of perturbation theory to compute the correct bispectrum and it should not be a surprise that the two differ. With that caveat in mind, we first expand the density field \eqref{eq:delta-Zel} to second order in the displacement:
\barr
\delta_{\rm Zel}(\bk) &\approx& \delta_{\rm L}(\bk) - \frac12 k^i k^j \int d^3 q ~ \rme^{i \bk \cdot \bq} \psi_i(\bq) \psi_j(\bq)\nonumber\\
&=& \delta_{\rm L}(\bk) + \frac12 \int \frac{d^3 k'}{(2 \pi)^3} \frac{\bk \cdot \bk'}{(\bk')^2} \frac{\bk \cdot (\bk - \bk')}{(\bk - \bk')^2}\nonumber\\
 && ~~~~~~~~~~~~~~~  \times  \delta_{\rm L}(\bk') \delta_{\rm L}(\bk - \bk'), \label{eq:delta_Zel_Q}
\earr
where in the second line we have used Eq.~\eqref{eq:psi-delta-fourier} to express $\psi_i$ in terms of the linear density field. 
The lowest-order bispectrum then follows in a straightforward manner:
\beq
B_{\rm Zel}(\bk_1, \bk_2, \bk_3) = F(\bk_1, \bk_2) P_{\rm L}(k_1) P_{\rm L} (k_2) + \textrm{ perm.}, \label{eq:B_Zel}
\eeq
where
\barr
F(\bk_1, \bk_2) &=& \frac{(\bk_1 + \bk_2) \cdot \bk_1 }{k_1^2} \frac{(\bk_1 + \bk_2) \cdot \bk_2 }{k_2^2} \nonumber\\
&=&  1 + \frac{\bk_1 \cdot \bk_2}{k_1 k_2} \left(
  \frac{k_1}{k_2} + \frac{k_2}{k_1}\right) + \frac{(\bk_1 \cdot \bk_2)^2}{k_1^2 k_2^2},~~~
\earr
is identical to equation (79) of Ref.~\cite{Bart3}, and different, as expected, from the correct result obtained in second order (Eulerian or Lagrangian) perturbation theory \citep{Bernardeau_2002}. 

Note that here we have expanded the density field first, for the ease of computation. One can also easily compute the bispectrum first, from the unperturbed expression for $\delta(\bk)$. When doing so one recovers the exponential damping prefactor of Ref.~\cite{Bart3}.

So far we have rederived part of the results of BFB, in the illustrative case of the Zeldovich approximation. We now turn to the most interesting and novel result, where the gravitational interaction is treated perturbatively.

\section{Perturbative treatment of the gravitational interaction}\label{sec:general}

\subsection{Trajectories of point particles in an expanding universe}

The starting point of our derivation are the equations
of motion of a non-relativistic point particle in an expanding Universe. We shall give a more formal (and exactly equivalent) description with the Vlasov-Poisson system in Section \ref{sec:Vlasov}. 

Denoting the scale factor by $a$, the conformal time by $\eta$ and the comoving
position by $\bs{x}$, the equations of
motion of a point particle are given by
\barr
\frac{d \bs{x}}{d \eta} &=& \frac{\bs{u}}{a}, \label{eq:xdot}\\
\frac{d \bs{u}}{d \eta} &=& - a \bs{\nabla} \phi. \label{eq:udot}
\earr
The Newtonian gravitational potential $\phi$ satisfies the Poisson equation:
\beq
\Delta \phi = 4 \pi G \overline{\rho} a^2 \delta = \omega_0^2 a^{-1} \delta,
\eeq
where $\delta \equiv \rho/\overline{\rho} - 1$ is the overdensity field, which has to be
determined from the trajectories themselves, and we have defined, for short,
\beq
\omega_0^2 \equiv 4 \pi G \overline{\rho}(z = 0) = \frac32 H_0^2 \Omega_m.
\eeq 
 The system
(\ref{eq:xdot})-(\ref{eq:udot}) has the following implicit integral solution:
\barr
\bs{x}(\eta) &=& \bs{x}(\eta_*) + (s - s_*) \bs{u}(\eta_*) \nonumber\\
 &-& \int_{\eta_*}^{\eta} (s - s')  a' \bs{\nabla} \phi(\bs{x}(\eta'), \eta')
d \eta', \label{eq:x-integral}
\earr
where $\eta_*$ is some initial time and 
\beq
s \equiv \int \frac{d \eta}{a}= \int \frac{dt}{a^2}
\eeq
is the ``superconformal time''. 

We assume that up to the initial time $\eta_*$, particle trajectories are well described by the Zeldovich approximation, so our initial conditions are: 
\barr
\bx(\eta_*) &=& \bq + \bs{\psi}(\bq, \eta_*), \\\label{eq:x-init}
\bs{u}(\eta_*) &=& a_* \frac{\dot{D}(\eta_*)}{D(\eta_*)} \bs{\psi}(\bq, \eta_*) \equiv \lambda_* \bs{\psi}(\bq, \eta_*), \label{eq:u-init}
\earr
with $\bs \nabla \cdot \bs{\psi}(\bq, \eta_*) = - \delta_{\rm L}(\bq, \eta_*)$. 
For compactness we shall define 
\beq
\alpha(\eta) \equiv 1 + \lambda_* (s - s_*).\label{eq:alpha}
\eeq
In what follows we shall write $\bs{\psi}(\bq, \eta_*) = \bs{\psi}(\bq)$, the dependence on $\eta_*$ being implicit.

\subsection{The gravitational interaction as a perturbation}

A standard technique in field theory is to treat interactions perturbatively. The new conceptual contribution of BFB is to use this technique for gravitational interactions in cosmology. In this context this approach is somewhat artificial as there is no explicit small parameter associated with the interaction -- in
fact, as we will show in the next section, even the simple linear fluid equations
resulting in the linear growth factor are formally of infinite order
in the interaction. Nevertheless, it should in principle converge to the
correct answer as higher and higher orders in the interaction are
included. It may moreover provide insights into the nonlinear
regime. 

In order to keep track of the order in the gravitational interaction, we insert a \emph{bookkeeping} perturbation parameter
$\epsilon$ in front of the potential. We shall set $\epsilon$ to unity at the end of the calculation.

Our starting point is therefore the following implicit solution for particle trajectories: 
\barr
\bs{x}(\eta) &=& \bs{q} + \alpha(\eta)
\bs{\psi}(\bs{q})\nonumber\\ 
&-& \epsilon \int_{\eta_*}^{\eta} (s - s') a' \bs{\nabla}
\phi(\bs{x}(\eta'), \eta') d \eta'.\label{eq:traj}
\earr
We shall expand the trajectories and density field to first order in $\epsilon$: 
\barr
\bs{x}(\eta) &=& \bx^{(0)}(\eta) + \epsilon ~\bx^{(1)}(\eta) + \mathcal{O}(\epsilon^2),\\
\delta(\bx) &=& \delta^{(0)}(\bx) + \epsilon ~ \delta^{(1)}(\bx) + \mathcal{O}(\epsilon^2).
\earr
The gravitational potential itself, being proportional to $\delta$ through the Poisson equation, is also a series in $\epsilon$. Since it is multiplied by $\epsilon$ in Eq.~\eqref{eq:traj}, we only need to use the zero-th order $\phi^{(0)}$ [sourced by $\delta^{(0)}$] when computing trajectories to first order in $\epsilon$. 

We therefore have, explicitly,
\beq
\bx^{(0)}(\eta) = \bq + \alpha(\eta) \bs\psi(\bq)
\eeq
for the free trajectory, and 
\beq
\bx^{(1)}(\eta) = - \int_{\eta_*}^{\eta} (s - s') a' \bs{\nabla}
\phi^{(0)}(\bx^{(0)}(\eta'), \eta') d \eta'\label{eq:x(1)}
\eeq
for the first-order correction. 

\subsection{Density field to first order in the interaction}
The density field is obtained by summing over all streams:
\beq
1 + \delta(\bx) = \int d^3 q ~ \delta_{\rm D}[\bx - \bx^{(0)}(\bq) - \epsilon~ \bx^{(1)}(\bq)].
\eeq
Expanding to first order in $\epsilon$ gives
\beq
1 + \delta^{(0)}(\bx) = \int d^3 q ~ \delta_{\rm D}[\bx - \bx^{(0)}(\bq)]. 
\eeq
and 
\beq
\delta^{(1)}(\bx) = - \int d^3 q ~ \bx^{(1)}(\bq) \cdot \bs{\nabla} \delta_{\rm D}[\bx - \bx^{(0)}(\bq)].\label{eq:delta(1)}
\eeq
Since the free trajectory $\bx^{(0)}= \bq+\alpha~ \bs{\psi}$ is, up to the coefficient $\alpha(\eta)$, identical to the Zeldovich trajectory, we immediately obtain $\delta^{(0)}(\bk)$ from equation \eqref{eq:delta-Zel}:
\beq
\delta^{(0)}(\bk, \eta) =  \int d^3 q ~ \rme^{i \bk \cdot \bq}\left[\rme^{ i \alpha(\eta) \bk \cdot \bs{\psi}(\bq)}  - 1\right]. \label{eq:delta(0)(k)}
\eeq

The Fourier transform of $\bs{\nabla} \delta_{\rm D}(\bs{x} -
\bs{x}_0)$ is just $-i \bs{k} \rme^{i \bk \cdot \bs{x}_0}$, and we thus get
\beq
\delta^{(1)}(\bk) = i \int d^3 q ~ \bk \cdot \bx^{(1)}(\bq) ~\rme^{i \bk \cdot \bx^{(0)}(\bq)}.\label{eq:delta(1)-int}
\eeq
From the Poisson equation, we get 
\beq
a \bs{\nabla} \phi^{(0)}(\bx, \eta) = i \omega_0^2 \int \frac{d^3 k'}{(2 \pi)^3} \frac{\bk'}{(k')^2} \delta^{(0)}(\bk', \eta) \rme^{-i \bk' \cdot \bx},
\eeq
so that Eq.~\eqref{eq:x(1)} becomes
\barr
\bx^{(1)}(\bq) &=& -i \omega_0^2 \int_{\eta_*}^{\eta} d \eta' (s - s') \int\frac{d^3 k'}{(2 \pi)^3} \nonumber\\
&& ~~~~  \frac{\bk'}{(k')^2} \delta^{(0)}(\bk', \eta') \rme^{-i \bk' \cdot \bx^{(0)}(\bq, \eta')}.
\earr
Inserting this into Eq.~\eqref{eq:delta(1)-int}, we arrive at
\beq
\delta^{(1)}(\bk, \eta) = \omega_0^2 \int_{\eta_*}^{\eta} d \eta' (s - s') \int\frac{d^3 k'}{(2 \pi)^3} \frac{\bk \cdot \bk'}{(k')^2} \mathcal{I}(\bk, \bk', \eta, \eta'),\label{eq:delta(1)(k)}
\eeq
where
\beq
\mathcal{I}(\bk, \bk', \eta, \eta') \equiv 
\int d^3 q ~ \delta^{(0)}(\bk', \eta') \rme^{i \bk \cdot \bx^{(0)}(\bq, \eta) - i \bk' \cdot \bx^{(0)}(\bq, \eta') }.
\eeq
Defining, for short, $\bs{K} \equiv \alpha(\eta) \bk$ and $\bs{K}' \equiv \alpha(\eta') \bk'$, we may rewrite this as 
\beq
\mathcal{I}(\bk, \bk', \eta, \eta') \equiv 
\int d^3 q ~ \delta^{(0)}(\bk', \eta')  \rme^{i (\bk - \bk') \cdot \bq } \rme^{i (\bs{K} - \bs{K}') \cdot \bs{\psi}(\bq)}.
\eeq
As a technical aside, note that $\delta^{(1)}(\bx)$ has zero mean, as can be seen by interchanging statistical averaging with averaging over the whole space in Eq.~\eqref{eq:delta(1)}. 

\subsection{Power spectrum to first order in the interaction}

We are now ready to compute the power spectrum. To first order in $\epsilon$, we have
\beq
P(k) = P^{(00)}(k) + 2 ~\epsilon~ P^{(01)}(k).\label{eq:P-int}
\eeq
Here $P^{(00)}$ is the power spectrum of $\delta^{(0)}$, which again is almost identical to the one in the Zeldovich approximation with the additional coefficient $\alpha^2$:
\barr
P^{(00)}(k) = \rme^{- k^2 \alpha^2 \sigma_{\psi}^2}\int d^3 q ~\rme^{i \bs k \cdot \bq} \left[\rme^{k^i k^j \alpha^2 C_{ij}(\bq)} -1 \right].~~~~ \label{eq:P00}
\earr
The first-order term comes from the cross-power of $\delta^{(0)}$ and $\delta^{(1)}$. It requires computing the following ensemble average:
\barr
\langle \mathcal{I}(\bk, \bk', \eta, \eta') \delta^{(0)*}(\bk'', \eta) \rangle\nonumber\\
= \int d^3 q ~ d^3 q' ~ d^3 q'' \rme^{i(\bk - \bk') \cdot \bq + i \bk' \cdot \bq' - i \bk'' \cdot \bq''}\nonumber\\
\times \Big{\langle} \rme^{i (\bs{K} - \bs{K}') \cdot \bs{\psi}(\bq) + i \bs{K}' \cdot \bs{\psi}(\bq') - i \bs{K''} \cdot \bs{\psi}(\bq'')}\Big{\rangle} ,
\earr
where $\bs{K}'' \equiv \alpha(\eta) \bk''$. Note that we got rid of the constant part of $\delta^{(0)}(\bk'')$, which is possible since $\delta^{(1)}$ has zero mean. We also got rid of the constant part of $\delta^{(0)}(\bk')$ since the Dirac function $\delta_{\rm D}(\bk')$ would vanish when multiplied by $\bk'$ in the final integral.

For any linear combination $A$ of gaussian random variables with zero mean\footnote{Short proof: if $A$ is a linear combination of gaussian variables, $A$ is itself a gaussian variable. Having zero mean, its moments are $\langle A^{2 p+1} \rangle = 0, \langle A^{2p} \rangle = (2 p)!/(2^p p!) \times \langle A^2 \rangle^p$. We then have $\langle \rme^A \rangle = \sum_p \langle A^{2p} \rangle/(2 p)! = \sum_p (\langle A^2 \rangle/2)^p /p! = \rme^{\langle A^2 \rangle/2}.$}, $\langle \rme^A \rangle = \rme^{\frac12 \langle A^2 \rangle}$. Working out the variance of the last term in brackets, we arrive at 
\barr
\big{\langle}... \big{\rangle} &=& \exp\Bigg{[}- \frac12 \sigma_{\psi}^2 \left[(\bs{K}' - \bs{K})^2 + (K')^2 + (K'')^2\right] \nonumber\\
&& ~~~~~~~ - ({K}^i - {K'}^i){K'}^j C_{ij}(\bq' - \bq) \nonumber\\
&& ~~~~~~~ +  ({K}^i - {K'}^i){K''}^j C_{ij}(\bq'' - \bq) \nonumber\\
&& ~~~~~~~ + {K'}^i {K''}^j C_{ij}(\bq'' - \bq') \Bigg{]}.
\earr
Changing integration variables to $\bq, \bq_1 \equiv \bq' - \bq$ and $\bq_2 \equiv \bq'' - \bq$, we see that the integral over $\bq$ leads to $(2 \pi)^3\delta_{\rm D}(\bk'' - \bk)$, as expected from statistical homogeneity. We then get
\barr
&&\langle \mathcal{I}(\bk, \bk', \eta, \eta') \delta^{(0)*}(\bk'', \eta) \rangle = \nonumber\\
&& P_{\mathcal{I}\delta^{(0)}}(\bk, \bk', \eta, \eta') (2 \pi)^3 \delta_{\rm D}(\bk'' - \bk),
\earr
with 
\barr
&&P_{\mathcal{I}\delta^{(0)}}(\bk, \bk', \eta, \eta') = \rme^{- \frac12 Q_{\rm D}} \int d^3 q_1 ~d^3 q_2 ~\rme^{i \bk' \cdot \bq_1 - i \bk \cdot \bq_2}\nonumber\\
&&\exp \Big{[}({K'}^i - {K}^i){K'}^j C_{ij}(\bq_1) +({K}^i - {K'}^i){K}^j C_{ij}(\bq_2) \nonumber\\
&& ~~~~~~~ + K^i {K'}^j C_{ij}(\bq_2 - \bq_1)  \Big{]}, \label{eq:PId0}
\earr
where we have defined, as in Ref.~\cite{Bart4},
\beq
Q_{\rm D} \equiv 2 \sigma_{\psi}^2 \left[{K'}^2 + K^2 - \bs{K} \cdot \bs{K}' \right].
\eeq
From this we finally obtain the contribution to the power spectrum at first order in the interaction
\barr
P^{(01)}(\bk, \eta) &=& \omega_0^2 \int_{\eta_*}^{\eta} d \eta' (s - s') \int\frac{d^3 k'}{(2 \pi)^3} \frac{\bk \cdot \bk'}{(k')^2}\nonumber\\
&& ~~~~~~~~~~~~~~ \times P_{\mathcal{I}\delta^{(0)}}(\bk, \bk', \eta, \eta').\label{eq:P01}
\earr
Equations \eqref{eq:P-int}, \eqref{eq:P00}, \eqref{eq:PId0} and \eqref{eq:P01}, along with the intermediate definitions, constitute one of our main results. They give the power spectrum to first order in the interaction, but \emph{to infinite order in the linear power spectrum}. We have indeed only expanded in $\epsilon$ and not in $P_{\rm L}$. This \revision{formal} result therefore goes beyond that of Ref.~\cite{Bart4}, which we shall now derive.

\subsection{Expansion in $P_{\rm L}$}

We now Taylor-expand our result in $P_{\rm L}(\eta_*)$ in order to recover the result of Ref.~\cite{Bart4}\footnote{The correspondence of notations is $g_{qp}(\tau, \tau') \equiv \lambda_* (s - s')$, $1 + g_{qp}(\tau, \tau_*) \equiv \alpha(\eta)$.}. The expansion of $P^{(00)}$ follows directly from our results for the Zeldovich approximation, equation \eqref{eq:Pzel-taylor} with an additional $\alpha^2$ multiplying $C_{ij}$ and $\sigma_{\psi}^2$. In particular, the linear part is
\beq
P^{(00)}_{\rm L}(k) = \alpha^2(\eta) P_{\rm L}(k, \eta_*). \label{eq:P00_L}
\eeq
This corresponds to equation (77) of Ref.~\cite{Bart4}.

We now expand $P^{(01)}$. The integrand of equation \eqref{eq:PId0} gets expanded into three linear terms and six quadratic terms. Only the terms that mix $\bq_1$ and $\bq_2$ survive integration, which leaves one linear term and four quadratic terms. The linear term is
\barr
&& P_{\mathcal{I}\delta^{(0)}, \rm L}(\bk, \bk', \eta, \eta') \nonumber\\
&=& \alpha \alpha'  (2 \pi)^3 \delta_{\rm D}(\bk' - \bk) k^i k^j \int d^3 q ~\rme^{- i \bk \cdot \bq} C_{ij}(\bq)\nonumber\\
&=& \alpha \alpha'   P_{\rm L}(k, \eta_*) (2 \pi)^3 \delta_{\rm D}(\bk' - \bk),
\earr
where we recall that $C_{ij}$ is the correlation tensor of the displacement field at $\eta_*$, hence the resulting $P_{\rm L}(\eta_*)$. The linear part of $P^{(01)}$ is therefore
\beq
P^{(01)}_{\rm L}(k, \eta) = \omega_0^2 \alpha(\eta) \int_{\eta_*}^{\eta} d \eta' (s - s') \alpha(\eta')  P_{\rm L}(k, \eta_*). \label{eq:P01_L}
\eeq
This is equivalent to equations (78)-(79)\footnote{Equation (79) of Ref.~\citep{Bart4} has a typographic error: the second argument of the two $g_{qp}$ should be $\tau_*$ instead of 0} of Ref.~\cite{Bart4}.

The four quadratic terms of the integrand of $P_{\mathcal{I} \delta^{(0)}}$ correspond to the terms $\overline{Z}^{(2A)}$ to $\overline{Z}^{(2D)}$ of Ref.~\cite{Bart4}, given in their equations (63) to (66). For example, the term involving the product of $C_{ij}(\bq_2)$ and $C_{ij}(\bq_2 - \bq_1)$ is two times
\barr
&&(K^i - {K'}^i)K^j  K^m {K'}^n \nonumber\\
&&\times \int d^3 q_1 ~ d^3 q_2 ~ \rme^{i \bk' \cdot \bq_1 - i \bk \cdot \bq_2}  C_{ij}(\bq_2) C_{mn}(\bq_2 - \bq_1)\nonumber\\
&& =  (K^i - {K'}^i)K^j  K^m {K'}^n \nonumber\\
&& \times \int d^3 q_1' ~ d^3 q_2 ~ \rme^{-i \bk' \cdot \bq_1' + i (\bk' - \bk) \cdot \bq_2}C_{ij}(\bq_2) C_{mn}(\bq_1')\nonumber\\
&& =  (K^i - {K'}^i)K^j  K^m {K'}^n \frac{({k'}^i - k^i)({k'}^j - k^j)}{|\bk' - \bk|^4} \frac{{k'}^m {k'}^n}{(k')^4} \nonumber\\
&& ~~~~ \times P_{\rm L}(\bk - \bk', \eta_*) P_{\rm L}(k', \eta_*) \nonumber\\
&& = \frac{\bs{K} \cdot \bs{K}'}{(\bk')^2} \frac{\bs{K} \cdot (\bk - \bk') (\bs{K} - \bs{K}') \cdot (\bk - \bk')}{|\bk' - \bk|^4}\nonumber\\
&&~~~~ \times P_{\rm L}(\bk - \bk', \eta_*) P_{\rm L}(k', \eta_*).
\earr
This is the term $\overline{Z}^{(2A)}$ of Ref.~\citep{Bart4}. Similarly, the term involving $C_{ij}(\bq_1)$ and $C_{ij}(\bq_2 - \bq_1)$ gives $\overline{Z}^{(2B)}$ and the term involving $C_{ij}(\bq_1)$ and $C_{ij}(\bq_2)$ gives $\overline{Z}^{(2C)}$. Finally, the square of $C_{ij}(\bq_2 - \bq_1)$ leads to 
\barr
&&K^i {K'}^j  K^m {K'}^n \nonumber\\
&& \times \int d^3 q_1 ~ d^3 q_2 ~ \rme^{i \bk' \cdot \bq_1 - i \bk \cdot \bq_2} C_{ij}(\bq_2- \bq_1) C_{mn}(\bq_2 - \bq_1)\nonumber\\
&& =  K^i {K'}^j  K^m {K'}^n (2 \pi)^3 \delta_{\rm D}(\bk' - \bk) \nonumber\\
&&~~~~~~~~~~~ \int d^3 q ~ \rme^{- i \bk \cdot \bq} C_{ij}(\bq) C_{mn}(\bq)\nonumber\\
&& =  (2 \pi)^3 \delta_{\rm D}(\bk' - \bk) \int \frac{d^3 k''}{(2 \pi)^3} P_{\rm L}(\bk - \bk'', \eta_*) P_{\rm L}(k'', \eta_*) \nonumber\\
&&~~~~ \frac{\bs{K} \cdot \bk'' ~ \bs{K}' \cdot \bk''}{(\bk'')^4} \frac{\bs{K} \cdot (\bk - \bk'') \bs{K}' \cdot (\bk - \bk'')}{|\bk - \bk''|^4}.
\earr
Up to a missing factor of $(2 \pi)^3$, this is the term $\overline{Z}^{(2 D)}$ of Ref.~\cite{Bart4}. Inserting this into equation \eqref{eq:P01} leads to equations (67) and (69) of Ref.~\cite{Bart4}.

Here again, as in the case of the Zeldovich approximation, Ref.~\cite{Bart4} approximate the term $\rme^{- Q_{\rm D}/2}$ by $(1 + Q_{\rm D}/2)^{-1}$, which they only apply to the quadratic term. As we pointed out previously, this approximation is not correct neither formally nor physically, as the term proportional to $- Q_{\rm D} P_{\rm L}$ is missing. In addition, the asymptotic behavior at large $k$ of the resulting power spectrum can be shown to be proportional to $P_{\rm L}(k)$. Again, this is due to the choice of the approximation for the damping prefactor, and the extending of this approximation much beyond its regime of validity.

In order to test the accuracy of the power spectrum to first order in the interaction against numerical simulations, one would need to numerically evaluate Eqs.~\eqref{eq:PId0} and \eqref{eq:P01}. This involves computing a triple three-dimensional integral, which is challenging, and is deferred to future work. 

\subsection{Going beyond the first order in the interaction: perturbed Vlasov-Poisson system} \label{sec:Vlasov}


We can extend our calculation to higher orders in the interaction. To do so, we may formulate the problem in terms of the Vlasov-Poisson system. In the relevant limit of an infinite number of particles, this is exactly equivalent to solving particle trajectories, but takes a more compact form for a formal description. 

The Vlasov-Poisson system is comprised of the collisionless Boltzmann equation for $f(\bx, \bs{u}, \eta)$, the phase-space distribution (divided by $\overline{\rho}(z = 0)$), and the Poisson equation:
\barr
\partial_{\eta} f + \frac{\bs{u}}{a} \cdot \partial_{\bx} f - \epsilon ~ a \bs{\nabla} \phi \cdot \partial_{\bs{u}} f = 0,\\
\Delta \phi = \omega_0^2 a^{-1} \int d^3 u ~ \delta f(\bx, \bs{u}, \eta),
\earr
where 
\beq
\delta f(\bx, \bs{u}, \eta) \equiv f(\bx, \bs{u}, \eta) -  \delta_{\rm D}(\bs{u})
\eeq
is the perturbation from a perfectly cold homogeneous matter. The initial conditions are 
\beq
f(\bx, \bs{u}, \eta_*) = \int d^3 q ~\delta_{\rm D}[\bx - \bq - \bs{\psi}(\bq)] ~ \delta_{\rm D}[\bs{u} - \lambda_* \bs{\psi}(\bq)].
\eeq
Expanding $f$ and $\phi$ in orders of the interaction, we have the following explicit recursion scheme:
\barr
\partial_{\eta} f^{(0)} + \frac{\bs{u}}{a} \cdot \partial_{\bx} f^{(0)} &=& 0,\\
\Delta \phi^{(n)} = \omega_0^2 a^{-1} \delta^{(n)} &=& \omega_0^2 a^{-1} \int d^3 u ~ \delta f^{(n)}(\bx, \bs{u}, \eta),~~\\
\partial_{\eta} f^{(n)} + \frac{\bs{u}}{a} \cdot \partial_{\bx}  f^{(n)} &=&   a \sum_{p = 0}^{n-1}\bs{\nabla} \phi^{(n-1-p)} \cdot \partial_{\bs{u}} f^{(p)}.
\earr
The inhomogeneous partial differential equation
\beq
\partial_{\eta} F + \frac{\bs{u}}{a} \cdot \partial_{\bx} F = S(\bx, \bs{u}, \eta)
\eeq
is most easily solved in Fourier space:
\barr
F(\bk, \bs{u}, \eta) &=& \rme^{i(s - s_*) \bs{u} \cdot \bk} F(\bk, \bs{u}, \eta_*) \nonumber\\
&+& \int_{\eta_*}^{\eta} d \eta' \rme^{i(s - s') \bs{u} \cdot \bk} S(\bk, \bs{u}, \eta').
\earr
The initial condition for the distribution function is, in Fourier space, 
\beq
f(\bk, \bs{u}, \eta_*) = \int d^3 q ~ \rme^{i \bk \cdot \bq + i \bk \cdot \bs{\psi}(\bq)} \delta_{\rm D}[\bs{u} - \lambda_* \bs{\psi}(\bq)],
\eeq
so that we have
\barr
&&f^{(0)}(\bk, \bs{u}, \eta) = \int d^3 q~ \rme^{i \bk \cdot \bq} \rme^{i \alpha \bk \cdot \bs{\psi}(\bq)} \delta_{\rm D}[\bs{u} - \lambda_* \bs{\psi}(\bs q)], ~~\\
&&f^{(n)}(\bk, \bs{u}, \eta) = -i\omega_0^2  \int_{\eta_*}^{\eta} d \eta' \rme^{i (s - s') \bs{u} \cdot \bk}\int \frac{d^3 k'}{(2 \pi)^3} \nonumber\\
&&~~~ \sum_{p = 0}^{n-1}\delta^{(n-1-p)}(\bk', \eta')\frac{\bk'}{(k')^2}\cdot \partial_{\bs u} f^{(p)}(\bk - \bk', \bs{u}, \eta').
\earr
The first-order calculation would recover the results we have obtained directly from particle trajectories. We see that going to higher orders in the interaction leads to a higher and higher number of terms, as well as a larger and larger number of nested integrals, if one is to keep the full nonlinear dependence on the initial conditions. This quickly becomes intractable for any practical computation.

A similar iterative solution to the Vlasov-Poisson system was suggested in Ref.~\cite{Tassev_2011}. There, the zero-th order solution is the phase-space density in the Zeldovich approximation, which accurately describes large scales, contrary to our zero-th order, non-interacting solution (as we shall see in the next section). The hierarchical solution of Ref.~\cite{Tassev_2011} is therefore potentially more accurate than an expansion in the gravitational interaction, but suffers from the same computational challenges when going to higher orders.

\section{Contact with standard linear perturbation theory}\label{sec:shortcomings}

The approach we have followed can allow us in principle to go to higher orders in the interaction, even though explicit computations would become more and more involved. It is best to understand beforehand what hope one may have of getting accurate results. This is best done by considering the power spectrum on large scales, where standard linear perturbation theory is very accurate. 

\revision{Since standard perturbation theory does not rely on an expansion in the gravitational interaction, it is, formally, \emph{of infinite order in the interaction}. In this section we shall study the convergence of the perturbative interaction treatment to standard linear perturbation theory on large scales.}

Since we are here concerned with linear theory (specifically, linear in $\delta_{\rm L}(\eta_*)$ or in the initial displacement field $\bs{\psi}$), we may treat the cold dark matter as an ideal fluid. The linearized continuity and momentum equations are
\barr
\frac{d \delta_{\rm L}}{d \eta} &=& - \theta_{\rm L}, \label{eq:fluid1}\\
\frac{d}{d \eta}(a \theta_{\rm L}) &=& - a ~ \epsilon~ \Delta \phi = -\epsilon~  \omega_0^2 \delta_{\rm L},\label{eq:fluid2}
\earr
where we have inserted the bookkeeping parameter $\epsilon$ in front of the gravitational potential and have used the Poisson equation in the second equality. This equation has an implicit integral solution:
\barr
\delta_{\rm L}(\eta) &=& \delta_{\rm L}(\eta_*) + a_* \dot{\delta}_{\rm L}(\eta_*) (s - s_*) \nonumber\\
&+& \epsilon ~ \omega_0^2 \int_{\eta_*}^{\eta} (s - s')  \delta_{\rm L}(\eta') d \eta'.\label{eq:growth-integral}
\earr
If we treat the gravitational interaction perturbatively, i.e. consider $\epsilon$ as a perturbation parameter, we first write
\beq
\delta_{\rm L} = \sum_n \epsilon^n \delta_{\rm L}^{(n)}. \label{eq:delta-series}
\eeq
The subsequent orders in the interaction can then be obtained from the
following recursion relation:
\barr
\delta_{\rm L}^{(0)}(\eta) &=& \delta_{\rm L}(\eta_*) + a_* \dot{\delta}_{\rm L}(\eta_*) (s -
s_*) \nonumber\\
&=& \alpha(\eta) \delta_{\rm L}(\eta_*), \\
\delta_{\rm L}^{(n)}(\eta) &=& \omega_0^2 \int_{\eta_*}^{\eta} (s - s')  \delta_{\rm L}^{(n-1)}(\eta') d \eta',\label{eq:delta(n)}
\earr
where $\alpha$ was introduced in Eq.~\eqref{eq:alpha}. 
The full solution is therefore
\barr
&&\frac{\delta_{\rm L}(\eta)}{\delta_{\rm L}(\eta_*)} = \alpha(\eta)+
\epsilon~ \omega_0^2 \int_{\eta_*}^{\eta} (s - s') \alpha(\eta')d
\eta' \nonumber\\
&& + ~
\epsilon^2 ~(\omega_0^2)^2 \int_{\eta_*}^{\eta} (s - s') d \eta'
\int_{\eta_*}^{\eta'} (s' - s'') \alpha(\eta'')d \eta'' + ... \nonumber\\
&&~~~~ = \mathcal{T}\left[ \rme^{\epsilon~
    \omega_0^2~\int_{\eta_*}^{\eta} (s - s') d \eta'}\alpha(\eta) \right], \label{eq:time-order}
\earr
where $\mathcal{T}$ is the time-ordering operator. 

We now recall that $\epsilon = 1$ is just a bookkeeping parameter. Solving the system of fluid equations \eqref{eq:fluid1}-\eqref{eq:fluid2} with $\epsilon = 1$ is equivalent to computing the full solution \eqref{eq:time-order} and leads to the usual linear growth factor $D(\eta)$:
\beq
\delta_{\rm L}(\eta) = \frac{D(\eta)}{D(\eta_*)} \delta_{\rm L}(\eta_*).
\eeq
On the other hand, truncating the series \eqref{eq:time-order} at any finite order in $\epsilon$ \emph{does not} recover the correct linear growth factor. 
Again, we emphasize that this is because standard linear perturbation theory accounts for the interaction to infinite order, even if it is, on the other hand, only linear in the initial conditions $\delta_{\rm L}(\eta_*)$. The two expansion schemes (in $\delta_{\rm L}(\eta_*)$ and $\epsilon$) are independent and this result is perfectly consistent.

We can now easily recover our previously obtained linear power spectrum at first order in the interaction:
\barr
&&P_{\rm L}(\eta) \approx P_{\rm L}^{(00)}(\eta) + 2 \epsilon P^{(01)}(\eta) \nonumber\\
&&= \Big{[}\alpha^2(\eta) +  2 \epsilon ~ \omega_0^2 \alpha(\eta)  \int_{\eta_*}^{\eta} d \eta' (s - s')  \alpha(\eta') \Big{]} P_{\rm L}(\eta_*), ~~~
\earr
which corresponds to equations \eqref{eq:P00_L} and \eqref{eq:P01_L} obtained in the previous section. 

\begin{figure}
\includegraphics[width = 85 mm]{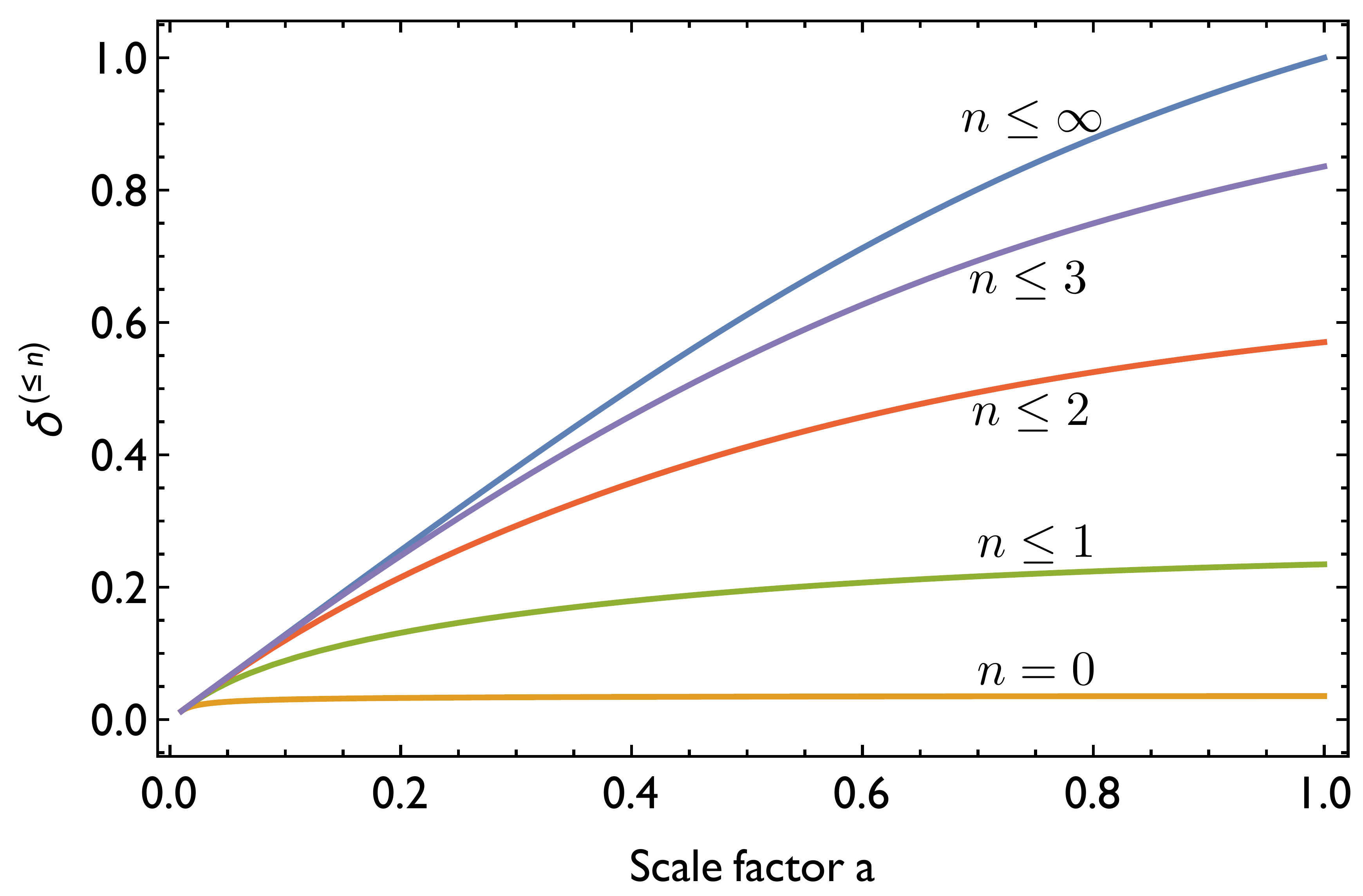}
\caption{Linear density field up to order $n$ in the
  interaction: for $n = 0, 1, 2, 3$ and $\infty$ from bottom to top. The normalization is such that the linear growth factor (corresponding to $n \leq \infty$) is unity
  at redshift zero. The starting conformal time $\eta_*$ corresponds to redshift
  $z_* = 99$. Cosmological parameters consistent with \emph{Planck} results are used.} \label{fig:deltas}
\end{figure}

Now how accurate an approximation do different orders in the interaction expansion provide to the correct linear growth factor? We first notice that for all $n$, $\delta^{(n)} \geq 0$, as can be seen from the recursion relation \eqref{eq:delta(n)}. Therefore any truncation of the series \eqref{eq:delta-series} will systematically underestimate the correct growth factor. We show in Figure \ref{fig:deltas} the linear growth factor summed up to order $n$ in the interaction, for $n = 0, 1, 2, 3$ and $\infty$, the latter case corresponding to the standard linear growth factor. Our starting time corresponds to redshift $z_* = 99$. We see that the series converges very slowly to the correct answer at redshift zero, and still grossly underestimates the correct linear growth factor even at third order in the interaction. 

To compensate for this missing power, Ref.~\cite{Bart4} only use the interaction expansion starting from a rather low redshift\footnote{Typically around redshift 10 or so \cite{Bart_private}.}. The interaction expansion is indeed more accurate closer to the starting redshift. The choice of the starting redshift is however arbitrary\footnote{In Ref.~\cite{Bart4}, this redshift $z_*$ is presumably determined by requiring that the evolution with the Zeldovich approximation from some early initial time $z_i$ to $z_*$, followed by the perturbative interaction approach from $z_*$ to redshift 0, result in the correct linear growth factor for large, linear scales. This can never be the case since, as we have shown, the perturbative interaction approach systematically underestimates the linear growth factor. It is therefore unclear how one should choose the starting time.}.

In closing of this section, we point out that BFB do recover the standard linear growth factor at large scales when they replace the hamiltonian propagator, correctly describing particles trajectories, by the ``Zeldovich propagator" (equation (65) of Ref.~\cite{Bart2}). This amounts to using the Zeldovich approximation. They do not derive the Zeldovich propagator nor the linear growth factor from first principles with the perturbative interaction approach. In this section we have explicitly shown how to do so, by going to infinite order in the interaction.

\section{Conclusions}\label{sec:concl}

Recently Bartelmann et al.~have suggested a new approach for computing statistics of CDM in the nonlinear regime. Their approach is inspired by field theory and relies on a formulation similar to the path integral. The main physical ingredient of this approach is to treat the gravitational interaction perturbatively, as one would do for other fundamental interactions in the weakly interacting regime. This approach is perfectly valid from a formal point of view and potentially very promising, calling for further examination. 

In this paper we have tackled two points, one regarding the technique and the second the essence of the approach.

First, we have concisely rederived all the results of BFB with a different formalism, relying on the integral solution of particle trajectories. We have furthermore derived the matter power spectrum to first order in the interaction but to infinite order in the linear power spectrum, going beyond the one-loop results of BFB. In addition, we have laid out the iterative equations one would have to solve to go to higher order in the interaction, using the Vlasov-Poisson system.

Second, we have made contact with standard linear perturbation theory and examined the validity of treating the gravitational interaction perturbatively. We have shown that standard linear perturbation theory accounts for the gravitational interaction at infinite order (even if it accounts for initial conditions at linear order). We have also shown that the convergence of a perturbative interaction approach is very slow on large, linear scales.

It is yet unclear whether such an approach would fare better in the nonlinear regime. The properties of the density field on highly nonlinear scales indeed depend on the details of bound or marginally bound structures, which may be difficult to reproduce if treating the gravitational interaction perturbatively. Perhaps the theory can produce accurate results if one applies it repeatedly over short enough intervals of time, though one then has to account for non-gaussian initial conditions. A rigorous study of these issues calls for a detailed analysis of the nonlinear regime using the interaction expansion, which we defer to future work.

\section*{Acknowledgments}
This work was supported by the John Templeton Foundation award 43770. 
I would like to thank Matthias Bartelmann, Liang Dai, Daniel Grin, Marc Kamionkowski, Ely Kovetz, Mark Neyrinck, Xin Wang and Matias Zaldarriaga for useful discussions and comments.

\bibliography{zeldovich.bib}

\end{document}